\documentclass[12pt,preprint]{aastex}

\newcommand{\rphk}{\ensuremath{R'_{\mbox{\scriptsize HK}}}}
\newcommand{\sdiff}{\ensuremath{S_{\mbox{\scriptsize diff}}}}
\newcommand{\caii}{\ion{Ca}{2} H \& K}

\shorttitle{Stellar Activity in 1231 Northern Hemisphere Stars}
\shortauthors{Wright, et al.}

\begin{document}

\title{Chromospheric \ion{Ca}{2} emission in Nearby F, G, K, and M stars\altaffilmark{1}}
\altaffiltext{1}{Based on observations obtained at Lick Observatory, which is
operated by the University of California, and on observations obtained
at the W. M. Keck Observatory, which is operated jointly by the
University of California and the California Institute of Technology.
The Keck Observatory was made possible by the generous financial support of the W.M. Keck Foundation.}

\author{J. T. Wright\altaffilmark{2}, G. W. Marcy\altaffilmark{2}, R. Paul Butler\altaffilmark{3}, S. S. Vogt\altaffilmark{4}}

\altaffiltext{2}{Department of Astronomy, University of California at
Berkeley, 601 Campbell Hall, Berkeley, CA 94720-3411;
jtwright@astro.berkeley.edu}

\altaffiltext{3}{Department of Terrestrial Magnetism, Carnegie
Institution of Washington, 5241 Broad Branch Road NW, Washington, DC 20015-1305}

\altaffiltext{4}{University of California / Lick Observatory, University of California at Santa Cruz, Santa Cruz, CA 95064}

\begin{abstract}
We present chromospheric \caii\ activity measurements, rotation
periods and ages for $\sim$1200
F-, G-, K-, and M- type main-sequence stars from
$\sim$18,000 archival spectra taken at Keck and Lick Observatories as
a part of the California and Carnegie Planet Search Project.  We have
calibrated our chromospheric S values against the Mount Wilson chromospheric
activity data.  From these measurements we have calculated median
activity levels and derived \rphk, stellar ages, and rotation periods
from general parameterizations for
1228 stars, $\sim$1000 of which have no previously published S
values.  We also present precise time series of activity measurements
for these stars.
\end{abstract}


\section{Introduction}
The California \& Carnegie Planet Search Program has included
 observations of $\sim$2000 late-type main-sequence stars at high
spectral resolution as the core of its ongoing survey of bright, nearby stars
to find extrasolar planets through precision radial velocity
measurements \citep[e.g.,][]{Cumming1999,Butler2003}.  One source of error in the measured velocities is that
due to ``photospheric jitter'':  flows and inhomogeneities on the
stellar surface can produce variations in the measured radial velocity
of a star and may even mimic the signature of planetary
companions (\citet{Henry2002,Queloz2001,Santos2003}).
In addition to providing a precision radial velocity,
each of our radial velocity observations provides a measurement of the strength of the stellar chromospheric
\caii\ emission cores.  These measurements are an indicator
of stellar magnetic activity and can provide an estimate of the
photospheric jitter and rotation period of a star, both critical
values for understanding and interpreting the noise present in radial
velocity measurements \citep{Noyes1984,Saar2000,Santos2000}.

The largest campaign to measure and monitor \caii\ emission has been the Mount
Wilson program begun by O. C. \citet{Wilson1968} and
continued and improved since then by \citet{Vaughan1978} and others \citep{Baliunas1998}.
From 1966-1977 this program used the ``HKP-1'' photometer which
employed a photoelectric scanner at the coud\'e focus of the 100 inch
telescope.  Since 1977 the ``HKP-2'' photometer has been used, which
is a new,
specially designed photomultiplier mounted at the Cassegrain focus of
the 60 inch telescope \citep[e.g.,][]{Baliunas1995}.

\citet{Duncan1991} published data from this program in the form of ``season averages'' of H \& K
line strengths from 65,263 observations of 1296 stars (of all
luminosity classes) in the Northern
Hemisphere, and
later as detailed analyses of 171,300 observations of 111 stars
characterizing the varieties and evolution of stellar activity in
dwarf stars.  This program defined the Mount Wilson ``S value'' which has become the standard
metric of chromospheric activity.

\citet{Henry1996} published data from a survey of stars in the southern
hemisphere, providing S values from 961 observations of 815 stars.
Other surveys include the Vienna-KPNO survey \citep{Strassmeier2000},
whose motivation was to find Doppler-imaging candidates by using \caii\ as a tracer of rotation period in 1058 late-type stars, and that of our
Anglo-Australian Planet Search \citep{Tinney2002}, which reported S
values for 59 planet search stars not observed by previous surveys.

\section{Observations}
Observations for the California and Carnegie Planet Search Program
have used the HIRES spectrometer at Keck Observatory for six years,
measuring precision velocities of 
$\sim$700 stars as part of a campaign to find and characterize
extrasolar planetary systems \citep[e.g.][]{Butler1996}.  HIRES is an
echelle spectrometer which yields high resolution 
(67,000) spectra from 3850\AA\ to 6200\AA.  Typical
exposures in the \caii\ region yield a signal
to noise ratio of 60 in the continuum, although this number can be smaller
for very red stars since our requisite signal to noise in the
iodine region of the spectrum dictates exposure time.

The detector on HIRES is a Tektronix 2048EB2 engineering-grade CCD
optimized for the optical.  The quantum efficiency degrades
significantly blueward of the H \& K lines, but is still 60\% at 0.38
microns.  Observations at Keck always employ an image rotator to keep
the position angle parallactic, thereby minimizing the effects of
atmospheric dispersion \citep{Vogt1994}.

The Planet Search program has also included observations made at Lick Observatory since
1987 with the Hamilton spectrograph fed by the Shane 3m telescope
and the 0.6-meter Coud\'e Auxiliary Telescope (CAT) \citep{Vogt1987}.  The
Hamilton spectrograph is also an echelle spectrometer with high
resolution (60,000).  In 2001 the CCD readout window was expanded to
include the \caii\ region, where we typically achieve a signal
to noise ratio between 10 and 60 in the continuum.  This large range of S/N is
partly due to the fact that the program does not employ an image
rotator at Lick in order to keep the overall system efficiency
high.  As a result significant blue flux can be lost for those
observations at large hour angles and airmasses \citep{Filippenko1982}.

Two detectors have been used on the Hamilton spectrograph since the
readout window was expanded.  The first CCD, referred to as ``Dewar 6'' for
the dewar it sits in, is an over-thinned 2048 $\times$ 2048 chip with
$15\mu$ pixels.  The second, ``Dewar 8'', is a Lawrence Berkeley
Laboratory high-resistivity CCD with the same dimensions as Dewar 6.

\section{The Stellar Sample\label{sample}}
The Planet Search at Lick and Keck observatories has included over
1000 stars over its duration, with many stars being added and a few dropped
along the way as resources and circumstances dictate.
\citet{Cumming1999} analyzed the frequency of planets around many (76)
of the best-observed Lick program stars, and \citet{Nidever2002} published absolute radial
velocities for most program stars at Lick and Keck and many other
stars (889) observed in the course of the program.  These same spectra
were analyzed by LTE atmosphere modeling to yield $T_{\mbox{eff}}$,
$\log{g}$, $v\sin{i}$, and chemical abundances by \citet{Valenti2004}
and \citet{Fischer2004}.  The sample of this paper
includes every star observed at Keck and Lick for which an accurate S can be
obtained.  It is composed of 1231 stars, 1199 of which have at least one
measured S from Keck and 132 of which have at least one from Lick.

The Planet Search initially included single, late type, dwarf stars
accessible from Lick Observatory.  As the resources of the Planet
Search have grown,
fainter F, G, and K dwarfs, M dwarfs, subdwarfs, and some subgiant
stars have been added, and as the effect
of activity on velocity precision has been uncovered, some more-active
stars have been dropped.  This paper's sample constitutes the stars
which were still being monitored when activity measurements began
(some now dropped)
well as stars added to the search since then and a few incidental
targets.  

\section{Data Reduction}
\subsection{S values}
The S value is defined by the operation of the Mount Wilson
spectrometers \citep{Duncan1991}, which measure a quotient of the
flux in two triangular bandpasses centered on the H \& K emission
cores and two continuum regions on either side.  \citet{Duncan1991} refer
to these channels as the H, K, R, and V channels (where the R and V
channels are the continuum channels on the red and blue sides,
respectively of the H and K channels.)  The HKP-1 spectrometer 
measured two 25 \AA-wide R and V channels separated by about 250
\AA\ about the rest position of the H \& K lines, and the H and K
channels which had a triangular instrumental profile with FWHM close to 1 \AA.
The HKP-2 spectrometer consisted of two 20 \AA-wide R and V channels
centered on 4001.07 \AA\ and 3901.07 \AA\ in the star's frame, and
triangular bandpass H and K channels with a
FWHM of 1.09 \AA.  Because the two Mount Wilson spectrometers had
different bandpasses defined for the four channels, Duncan et
al. derived a transformation from the HKP-1 measurements (referred to
as ``F values'') to the HKP-2 S values.

The S values were constructed as 
\begin{equation}
S = \alpha \frac{H + K}{R + V}
\end{equation}
where $H$, $K$, $R$, and $V$ refer to the flux in the corresponding
bandpasses, and $\alpha$ was calculated to be 2.4 to make the mean S
correspond to the mean F determined from the HKP-1
observations.

The differences in the continuum regions resulted in a transformation
being necessary from S to F:
\begin{equation}
F=0.033 + 0.9978 S - 0.2019 S^2
\end{equation}

We follow a similar prescription to extract measurements of activity
from our spectra and transform those values into S values on the Mount
Wilson scale.

\subsection{Reduction of Spectra and Calibration of S\label{reduction}}
\subsubsection{The Planet Search Reduction Pipeline}
For all Planet Search spectra, extraction from raw CCD
images is performed in an automated pipeline.  We 
measured S values from these archival, reduced spectra.  

We apply a scattered light subtraction to the HIRES echellograms
before extraction.  HIRES echellograms have many pixels of inter-order
real estate, from which we can make good measurements of scattered
light.  We fit B-splines to the signal in these inter-order regions
and interpolate linearly between them to estimate the scattered light
in each order. 

For both Lick and Keck data, a cosmic-ray removal algorithm removes
the strongest cosmic rays from each 2-D echellogram before
extraction. This is performed by modeling the profile of each order in
the spatial direction by averaging over a suitably large region in the
wavelength direction.  Having determined the spatial profile in a
region of the echellogram, cosmic rays are identified as extreme
excursions from the mean profile.  This technique is nearly identical
to the technique used in optimal extraction \citep[][ \S\S II {\sc d}
\& {\sc e}]{Horne1986}.   

One difficulty in extracting S values from Planet Search program spectra
is that the spectra are not flux calibrated.  Further, to properly
account for photon statistics, the blaze function and throughput of the
spectrometers are not removed.

\subsubsection{Extraction of $S$ from Keck Observations}
We extracted S values from our Keck spectra following the prescription
of \citet{Duncan1991} as closely as possible.  To
remove the blaze function, which the standard extraction retains, as
noted above, we smoothed and normalized a representative
flat field spectrum and divided it out of all spectra in the region of
interest. This replaced the blaze function, which is a strong function of
position along each spectral order, with the much more slowly varying
continuum of the quartz lamp used in our flat field images.  

We simulated the measurement of the Mount Wilson spectrometers by summing the
counts within four wavelength bins.  We defined two triangular, 1.09 \AA\
FWHM bandpasses centered on the H \& K lines, and two, fixed continuum
channels 20 \AA\ wide, just as in the case of the HKP-2 spectrometer.
One difference we employed was to fix the continuum regions in the
observer's frame rather than shift them into the star's frame.  This
prevented stars with particularly large Doppler velocities from
shifting the channel into the next order, and it allowed us to correct
for the effects of the imperfect flux calibration, as discussed
below.  The effect of choosing this frame, rather than the stellar
frame, was extremely small, causing changes in S of less than 1\%. 

\begin{figure}
\plotone{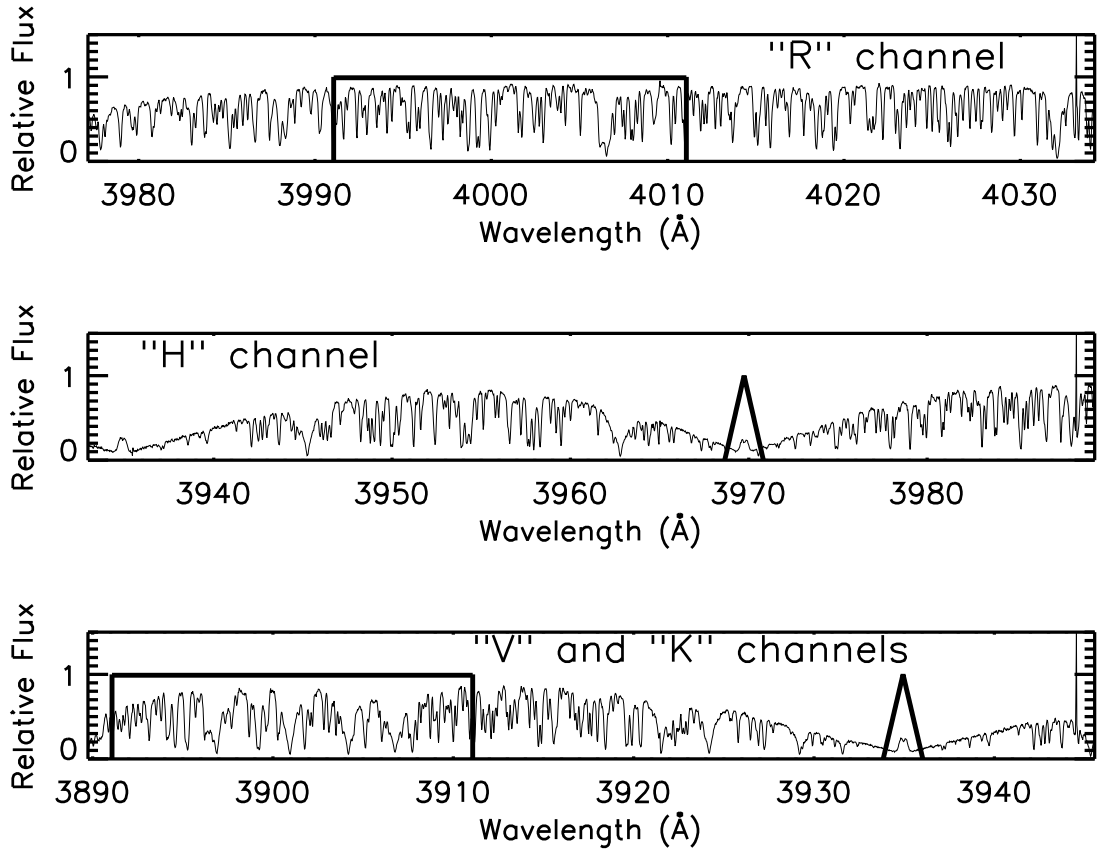}
\caption{R, H, K, and V channels in a representative Keck spectrum.  The
ordinate is relative photon flux in arbitrary units.  Wavelength is in
the rest frame of the star.  The H and K channels are always centered
on the line cores;  the R and V channels are fixed in the observer's
frame.\label{fig1}}
\end{figure}

Figure~\ref{fig1} shows the positions of the R, H, K, and V
channels in one of our Keck spectra.  Because we do not flux calibrate
our spectra the relative fluxes of neighboring orders is not correct
and the small effect of dividing out a quartz-lamp flat field remains.
Because our continuum channels are fixed in the observer's frame,
these spurious effects are essentially a constant function of the
spectrometer, not of the object being observed.  As a result the
relative fluxes calculated by integrating the fluxes within each of
the four bandpasses is different from the proper value by a constant
multiplicative factor. 

Rather than model and calculate these factors we elected to fit for
them and solve for any additional calibrations required to match
Mount Wilson S values without invoking additional degrees of freedom.
We thus constructed S values from our Keck spectra as:
\begin{equation}
S=\frac{a H + b K}{c R + d V}
\end{equation}
where $a$, $b$, $c$, and $d$ are relative weights to be determined.
We found that there are 114 suitable stars in our sample
which have S values published in Duncan et al., and which have been observed
more than once by each of our projects.  We used unweighted
averages of the Mount Wilson seasonal data to construct a mean S
value, and performed a non-linear least-squares fit in log space for
our relative  weights.  We found a good solution over all spectral
types:
\begin{equation}
S=\frac{1.68 H + 0.585 K}{0.497 R + 1.72 V}
\end{equation}
The addition of constant or $B-V$ terms did not improve $\chi_\nu^2$ of
our fit.

Not all of our extracted spectra were of sufficient quality for our
purposes.  The automated extraction pipeline occasionally failed to
properly trace an order on the echellogram or it improperly extracted
the background scattered light.  We found that an excellent diagnostic of the quality of a spectra was
to simply examine the ratio of the counts in the H and K channels.  We
modeled the dependence of the H/K ratio as a function of derived S
with a spline, and rejected all points for which the H/K ratio differed from
this model curve by a factor of 1.35 or more.  

A second rejection was based on the empirical observation that a few
very low-S/N spectra passed the ratio test but were clearly useless.  We rejected all stars with fewer than 200 counts per
pixel in the V channel.  These cuts culled 759 of 15,274 spectra,
bringing to total remaining Keck spectra to 14,515.

\begin{figure}
\plotone{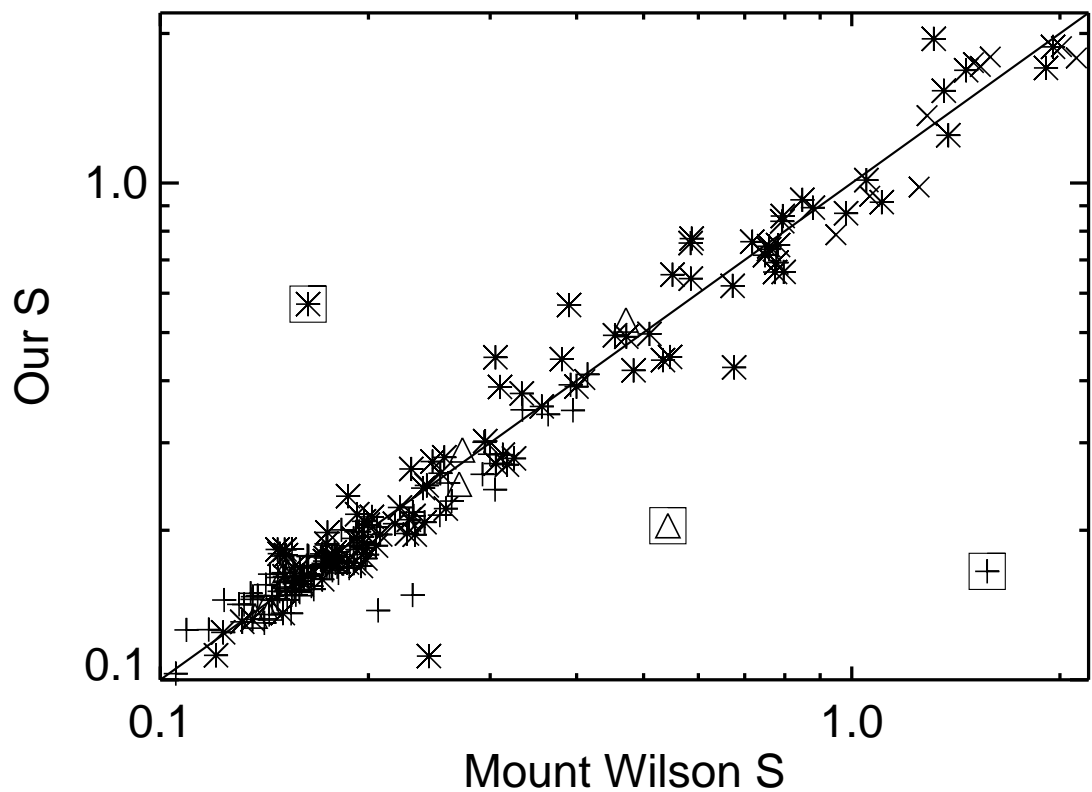}
\caption{Median Keck S from this work versus mean Mount Wilson S from
Duncan et al.  The scatter of 13\% (excluding the boxed outliers) is due in large part to the intrinsic chromospheric
variability of these stars.  Triangles represent F stars, crosses G stars, asterisks
K stars, and X's M stars.  We discuss the boxed outliers in
\S~\ref{targetnotes}.\label{keckcalib}}
\end{figure}

The results of the data rejection and calibration are shown in
Figure~\ref{keckcalib}, which shows Median S from Keck versus mean S
from Duncan et al. for the 199 stars our programs have in common.  The
13\% scatter in Figure~\ref{keckcalib} is primarily due to the
chromospheric variability intrinsic
to the stars.  The Mount Wilson measurements used for the Keck
calibration were made between 1966 and 1983 and many of these stars
simply have different levels of chromospheric activity today than
their means over that period.

\subsubsection{Extraction of $S$ from Lick Observations}
We calibrated Dewars 6 and 8 from Lick independently because different
extraction programs were used for the two dewars.  Data reduction for
our Lick spectra proved more difficult than it was for
the Keck spectra.  One complication is the lack of use of an image rotator at
Lick:  differential refraction can impose a spurious slope on the flux
spectrum which is impossible to calculate a priori and difficult to
calibrate. Also, since most of the Lick spectra are from the CAT, the
typical signal to noise is considerably lower than for the Keck spectra.

Dewar 8 spectra posed additional difficulties as well. The automated
extraction pipeline and observational plan for our Lick spectra were not
optimized for the blue orders, where no Doppler information is
gathered.  This causes difficulty because the orders on the
Hamilton spectrograph are closely spaced, making the extraction
algorithm sensitive to any error in the assumed position of the orders
which might arise from the low signal.   The Dewar 8 reduction algorithm
seems to have suffered along these lines, causing the continuum
normalization to be uncertain by 10\%.  Further, the high-resistivity
CCD in Dewar 8 yields a high incidence of cosmic-ray-like 
``worms'' which are difficult to remove and can contribute a
significant fraction of the flux in the H \& K and continuum channels.
As a result the Dewar 8 S values are less precise than those of the
other instruments. 

\begin{figure}
\plotone{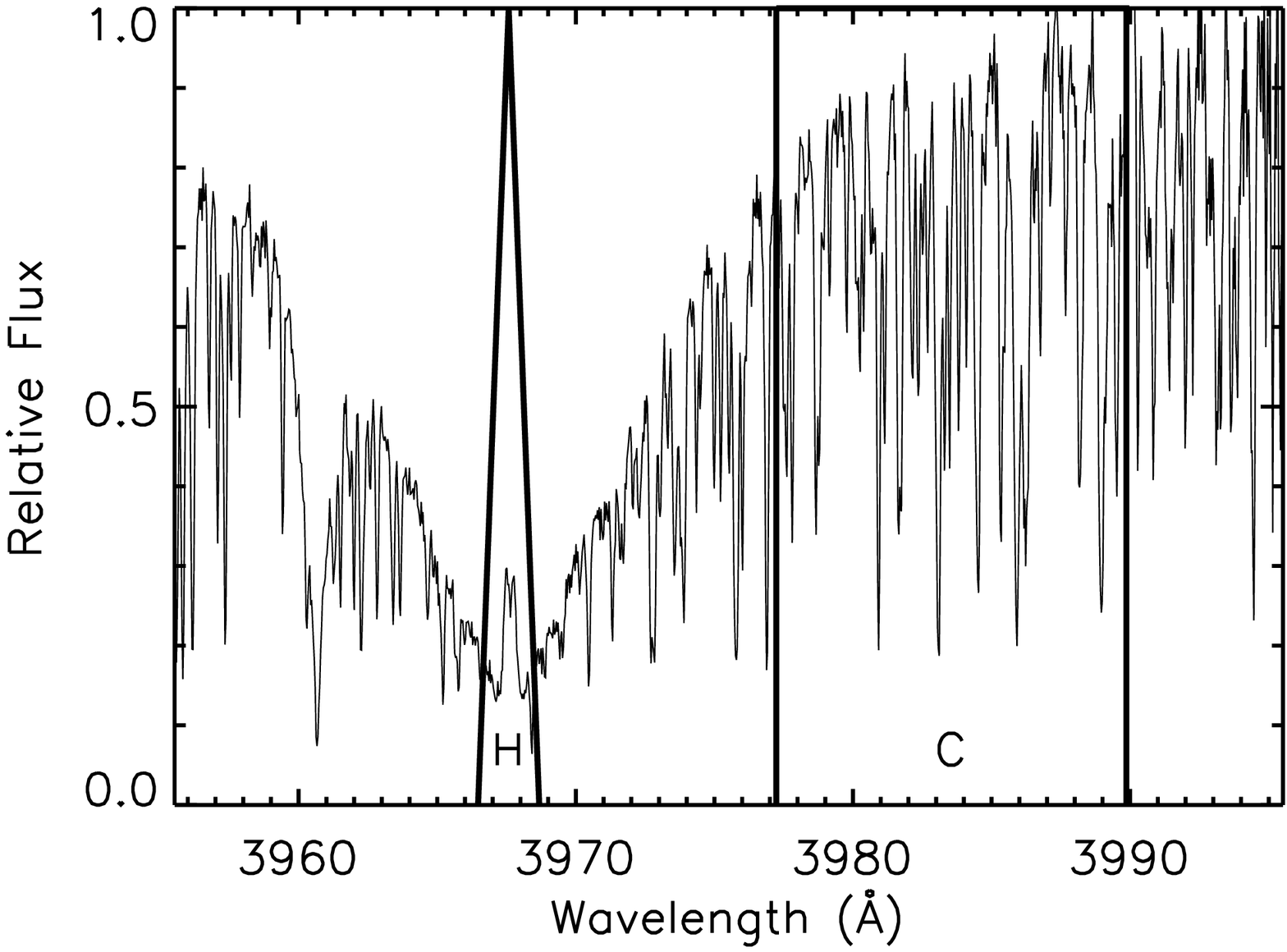}
\caption{The H and C channels for a representative Lick spectrum.  The
ordinate units are arbitrary, the abscissa is in Angstroms.\label{fig3}}
\end{figure}

To mitigate these problems we employed a simpler extraction algorithm
for our Lick spectra.  We simply defined a continuum region which we
denote the ``C'' channel just redward of the H line and constructed a
simple ratio of the H channel to the C channel (see Figure~\ref{fig3}).  
We thus defined our raw, uncalibrated Lick ``S value'' as
\begin{equation}
L=\frac{H}{C}
\end{equation}
where we use the symbol ``L'' to denote the fact that this is
not a Mount Wilson S value, but a ratio we have constructed related to
it.  The proximity of the continuum region and its employment of fewer
pixels reduced the severity of the problems outlined above.  

As a check, we examined the dependence of the extracted L on the
signal to noise in the H \& K region for a
well-observed star ($\tau$ Ceti).  The L values for exposures with low
signal were clearly discrepant from nominal values,
which is probably a result of poor extraction or poor accounting of
the flux zero point.  Such problems are not unexpected, since the
extraction and background subtraction algorithm was designed for and tested
on the typically high-signal iodine region.  To correct for this we
rejected all data in the low signal regime which we defined as spectra
with a signal to noise ratio of $<50$ in the continuum.  The effect of this strict
rejection scheme is severe.  Of 3014 spectra, only 1400 survived this cut.

\begin{figure}
\plotone{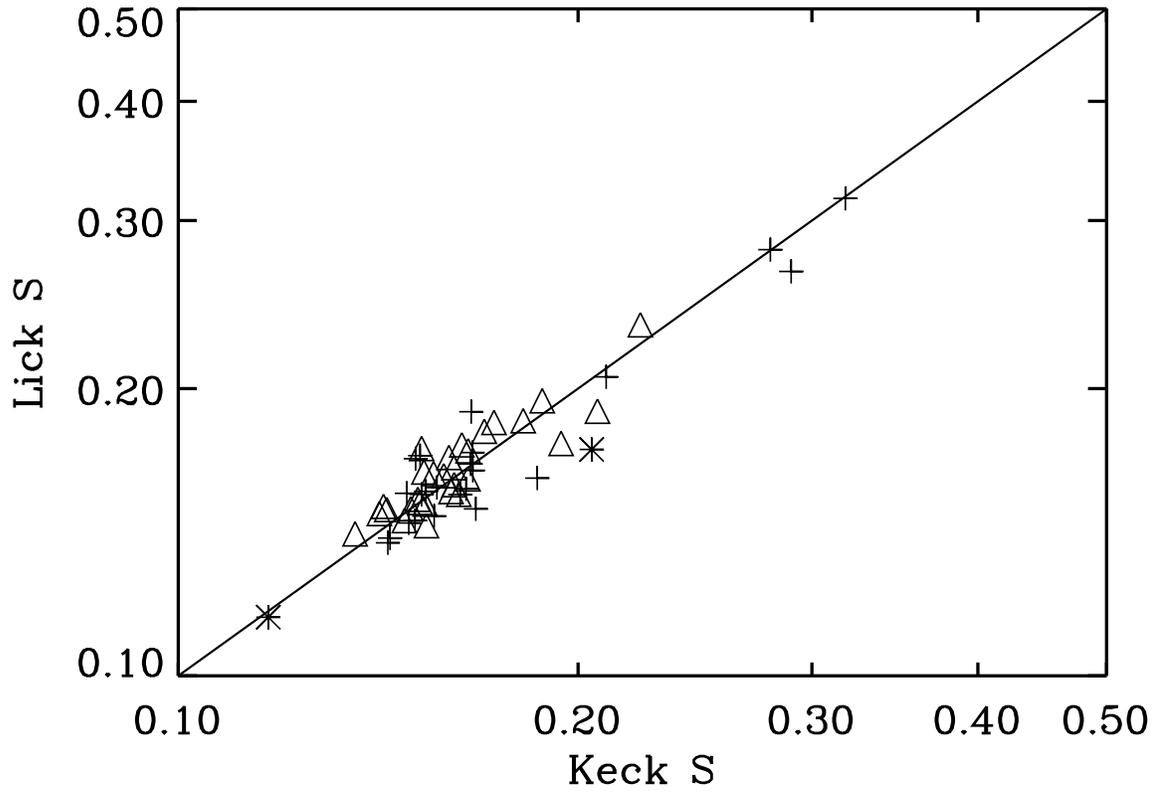}
\caption{S measurements from Dewar 6 at Lick versus Keck S.  The rms scatter is
6\%, some of which is 
due to the intrinsic variability of these stars.  Triangles represent
F stars, crosses G stars, and asterisks K stars.\label{d6calib}}  
\end{figure}

\begin{figure}
\plotone{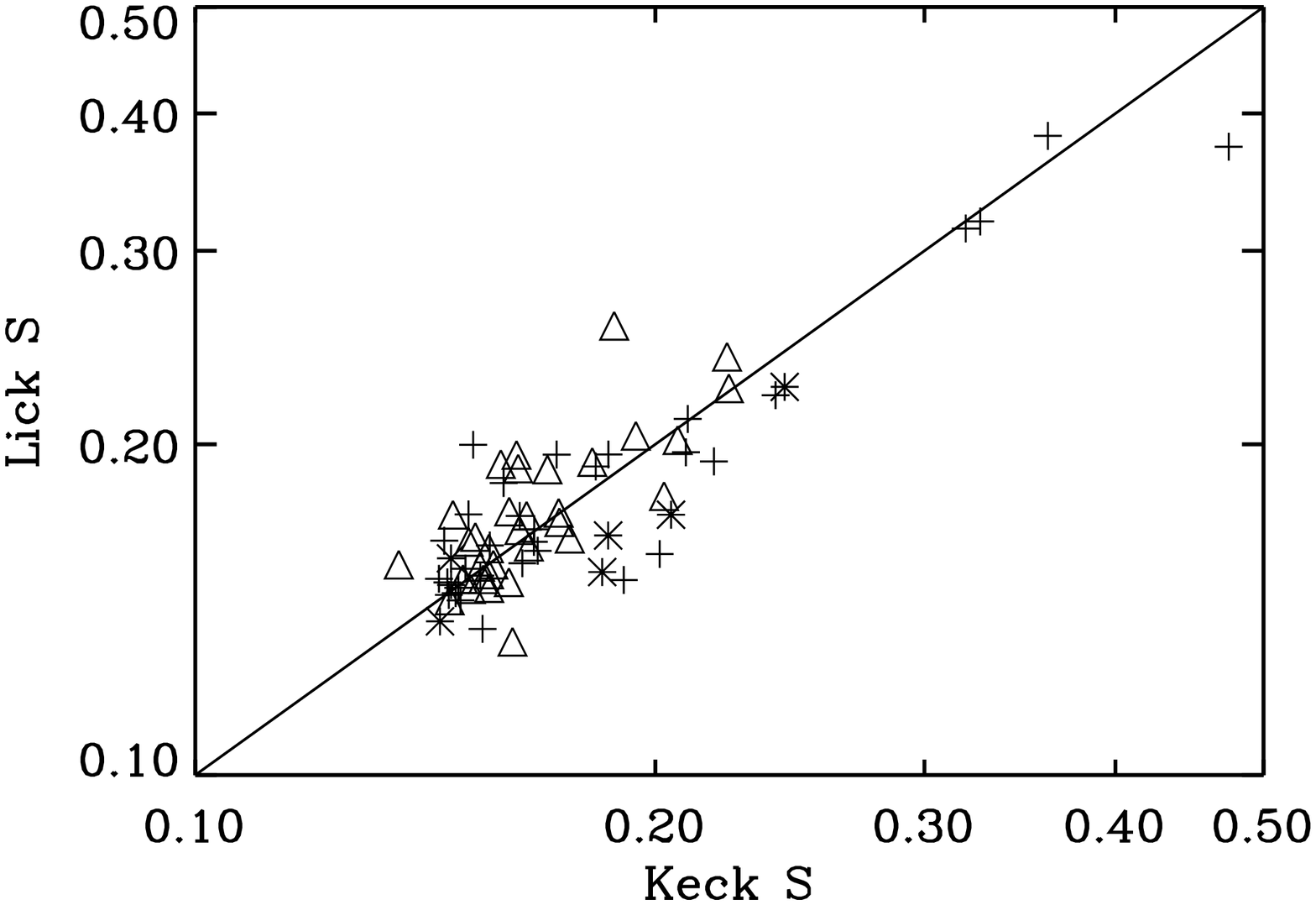}
\caption{S measurements from Dewar 8 at Lick versus Keck S.  The rms scatter is
12\%, some of which due to the intrinsic variability of these stars.  Triangles represent
F stars, crosses G stars, asterisks K stars.\label{d8calib}}
\end{figure}

Because there are very few stars in our Lick sample that have never
been observed at Keck, we performed a secondary calibration against
our Keck data.  We found that to match the Keck S values required quadratic
transformations.  For Dewar 6:
\begin{equation}
S = 0.507 L + 0.189 L^2
\end{equation}
and for Dewar 8:
\begin{equation}
S = 0.607 L + 0.239 L^2.
\end{equation}
Figures~\ref{d6calib} and \ref{d8calib} show how the final Lick S values
compare to the Keck measurements.  The scatter in Figure~\ref{d6calib} is
considerably lower than that of Figure~\ref{keckcalib} because the
measurements were contemporaneous.  

\subsection{Differential S values\label{differential}}
It is possible to make measurements of the flux change in an emission
line of a particular star which are much more precise than the
absolute strength indicated by $S$, as measured above.  Since changes in
$S$ can be important diagnostics of rotation and activity, we
have also measured sensitive differential S values,
\sdiff, for all of our Keck observations, inspired by the technique of
\citet{Shkolnik2003}.  We made these measurements from the same
reduced data as the S values measured above, but with an independent
technique, as described below.

For each star, we chose the observation from Keck with the
highest signal to noise ratio as a template and reference spectrum.
We scaled and shifted all other Keck spectra for that star 
such that we could directly compare the H line of every observation to
the reference spectrum.  When necessary, we added a small constant
offset or slope to each spectrum to match the reference spectrum as
closely as possible.  We then defined $E_i$, the summed (scaled)
counts in a 1 \AA\ rectangular bandpass in observation $i$.

\begin{figure}
\plotone{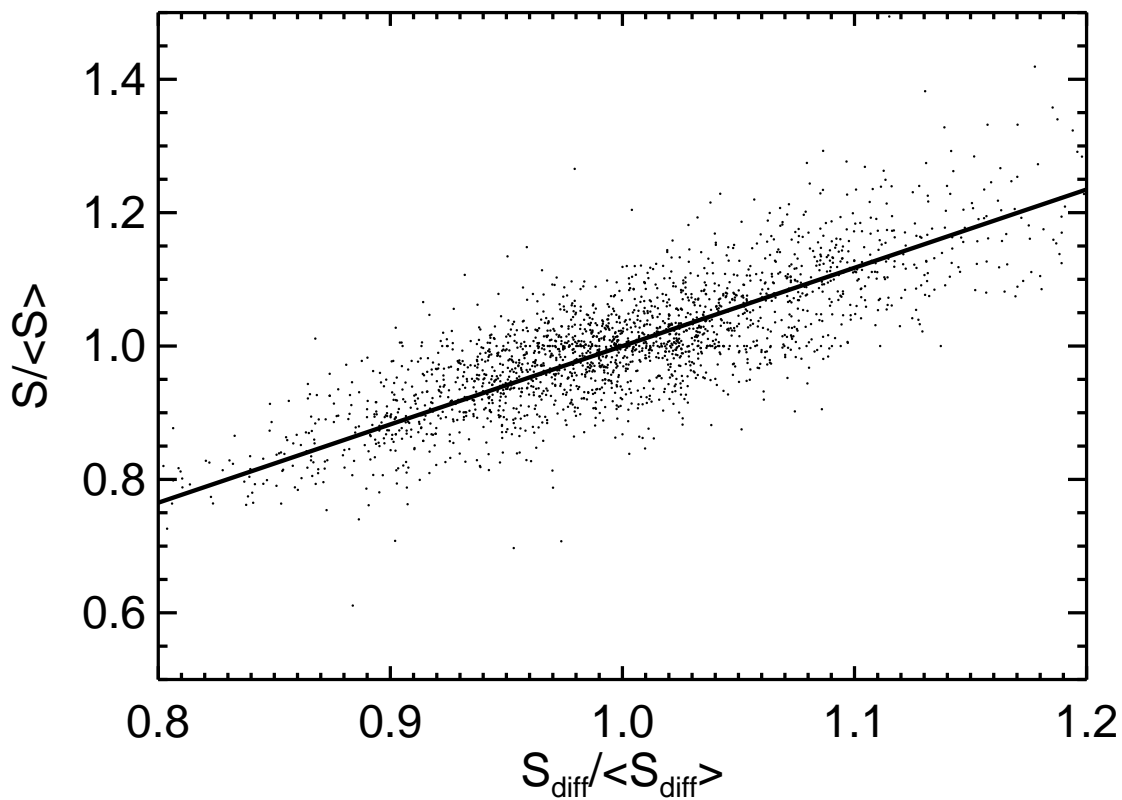}
\caption{This figure demonstrates the calibration of the
differential S values (\sdiff) to the scale of the less precise, absolute
measurements of $S$.  Each point
shows the $S$ and \sdiff\ measurements from a single Keck spectrum,
scaled by the median $S$ and mean \sdiff, respectively, for all such observations of a
given  star.  2203 observations of the 124 most chromospherically
variable stars are represented here.  The best-fit line is shown as a
solid line and has a slope of 1.2.  The 6\% scatter of this
distribution about the solid line is consistent with the estimate of
the errors in $S$ in \S~\ref{season}. \label{sdiffplot}}
\end{figure}

To transform $E$ to the same scale as the S values measured in
\S~\ref{reduction} we compared the fractional changes with time in $E$,
$\frac{E_i}{\bar{E}}$, to that in $S$, $\frac{S_i}{\bar{S}}$, for those stars with more than 9 Keck spectra and
more than 2\% variation in E, as shown in Figure~\ref{sdiffplot}.
We fit a single line to these points for all such stars using a robust line-fitting routine
which iteratively rejects outliers using the more precise \sdiff\
values as the independent variable. The best-fit line which passes
through (1,1) has a slope of $1.17$.  Therefore we adopt the relation:

\begin{equation}
S_{{\mbox{\scriptsize diff,}} i} = (\frac{E_i}{\bar{E}})^{1.2}\bar{S}
\end{equation}

where $\bar{E}$ is the mean value of $E_i$ for the star, and $\bar{S}$
is the grand $S$ for the star (see Section~\ref{tablelabel}).

\section{Data From H\& K Measurements}
\subsection{Uncertanties\label{errors}}
Estimating uncertainties in our final values of \rphk\ or other
chromospheric-based quantities is difficult because they vary
intrinsically with time on all timescales including those for
rotation and stellar activity cycles.  Activity cycles with
periods longer than the duration of our observations will not be well measured;
our final \rphk\ values for our sample therefore represent median activity
levels during the time we observed them, and not true averages for the star.

Measurement errors in these \rphk\ vaules stem from the modest signal to noise
ratio in the spectra and the quality of the calibration to the Mount
Wilson S value.  We
discuss below uncertainties from random errors from finite signal to noise ratios and short
term stellar variability.  Calibration errors are negligible due
to the large number of stars we used in the calibration.  

The 13\% scatter in Figure~\ref{keckcalib} is partly due to stellar
variability, since our data are not contemporaneous with the Mount
Wilson data and many stars are in different parts of their activity
cycles.  Thus our quoted values of R'HK carry uncertainties of no more than
13\%, that is, they lie within 13\% of the longterm average for the
typical star.  Measurements for
stars observed more frequently and for the full duration of the Planet
Search program will have correspondingly lower uncertainties.

\subsubsection{Random Errors\label{season}}
To estimate the random errors in our S values we used
$\tau$ Ceti (= HD10700, HR509, $V =$ 3.5) as a test case.  $\tau$ Ceti
serves as an excellent diagnostic star because we have a large number of
observations of it at Keck and with both dewars at Lick.  We observe
$\tau$ Ceti often because of the extraordinary velocity precision we
can achieve for this star with short exposure times.  This makes it an
excellent source with which to search for any systematic errors in our
precision velocities.  

$\tau$ Ceti is also very well observed by the Mount Wilson project.
\citet{Baliunas1995} note that despite its late spectral type and
color, it exhibits only 1\% variation in its S values, suggesting that
it may be in a ``Maunder Minimum''.  They also note that the
rotation period (33 d) implied by its S value and the observed $v \sin{i}$
(1 km/s) suggest that the star may be viewed nearly pole on.

\begin{figure}
\plotone{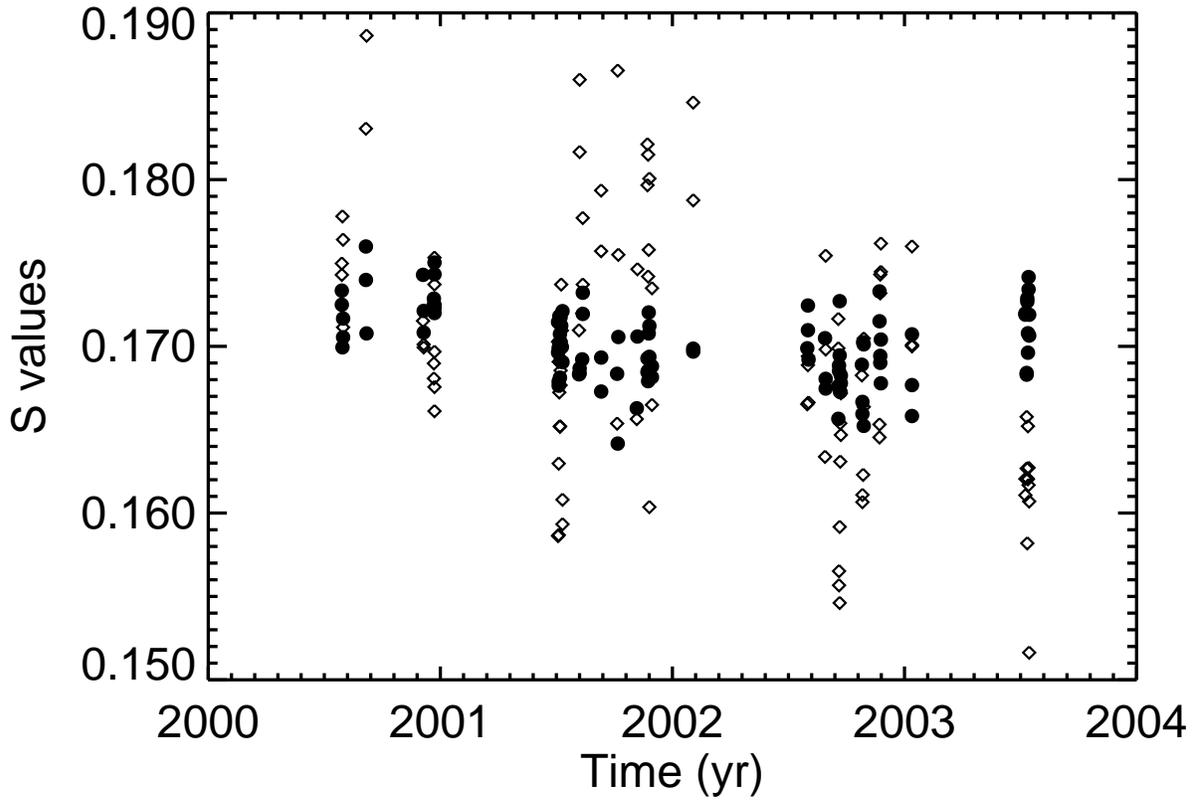}
\caption{$\tau$ Ceti S values (open diamonds) and differential S
values, \sdiff\ (filled circles) from Keck.  Activity
variations apparent in the S values are revealed to be due to uncertainties by
the more precise differential values.\label{taucetidiff}}
\end{figure}

The standard deviation of all $\tau$ Ceti S values is 6\% for the Keck
observations, 5.5\% for Dewar 6, and 10.5\% for Dewar 8.  The
differential S values from Keck (\S~\ref{differential}) for this star
(Figure~\ref{taucetidiff}) have a standard deviation of 1.3\%, which
is consistent with the 1\% variations reported by the Mount Wilson
project.  These estimates are consistent with the 6.5\% scatter about
the fit in Figure~\ref{sdiffplot}.
  
The disparity in the scatter in $S$ and \sdiff\ 
quantifies the errors introduced by difficulties in the raw reduction
and extraction of S values from the echellograms, as described in
\S~\ref{reduction}.  These include the difficulty of properly
correcting for scattered light, properly removing the blaze function
of the spectrometer, and properly extracting orders with modest signal
to noise.  Much of the systematic component of these errors is
removed by our calibration procedure, but a component which varies
with time clearly remains.  While the errors induced by these
difficulties are not necessarily characterized by Gaussian noise,
Figure~\ref{taucetidiff} shows that they are averaged out over many observations. 

Errors in $S$ for other stars will be similar to those for $\tau$ Ceti
because we use an exposure meter to ensure a uniform signal to noise
across our sample.  This exposure meter is sensitive to light in the
iodine region, so blue stars will have smaller Poisson errors in the
\caii\ region than red stars.  

We can calculate uncertainties in \sdiff\ another way, as well.
Occasionally during the course of the planet search, we take
two or more consecutive exposures of a star.  Under the assumption
that the chromospheric activity of the star does not change over the
course of several minutes, we can look at these sets of exposures and use the
variation in the \sdiff\ values measured in these sets as an estimate
of the precision of \sdiff.  Based on these sets we estimate a
typical uncertainty in \sdiff\ of 1.2\%.  This value, which is significantly higher than the
Poisson noise, represents a reasonable estimate of the precision of
the \sdiff\ values.  This value is also consistent the smallest
variations in \sdiff\ seen among stars in our sample. 

There are also a small systematic (and therefore correlated)
errors in the the differential S
values, on the order of 1\%.  For instance, our values of \sdiff\ have
a small and complex dependence on the focus of the
spectrometer.  This is can be seen in Figure~\ref{taucetidiff} where
there appears to be a slight decrease in the \sdiff\ values for $\tau$
Ceti after 2001 due to an improvement in the focus of HIRES at that
time.

\subsection{\rphk , Ages, and Rotation Periods\label{derived}}
The S index includes both chromospheric and photospheric
contributions.  To remove the photospheric component and determine
the fraction of a star's luminosity that is in the \caii\ lines,
we follow the prescription of \citet{Noyes1984} to generate the $\log{\rphk}$ values which appear in Table 2.  The transformation of S
indices to \rphk\ is a function of $B-V$, and is only calibrated
for $0.44<B-V<0.9$.

The transformation used by Noyes et al. is that of \citet{Middelkoop1982}:
\begin{equation}
R_{\mbox{\scriptsize HK}}=1.340 \times 10^{-4}
C_{\mbox{\scriptsize cf}} S
\end{equation}
where 
\begin{equation}
C_{\mbox{cf}}(B-V)=
1.13(B-V)^3 - 
3.91(B-V)^2 +
2.84(B-V)   -
0.47
\end{equation}
transforms the flux in the R and V channels to total continuum flux and S is the Mount Wilson S value of the star.  This number must be
corrected for the photospheric contribution to the flux in the \caii\
line cores.  Noyes et al. use the expression in \citet{Hartmann1984} 
\begin{equation}
\log{R_{\mbox{\scriptsize phot}}} = -4.898
           +1.918(B-V)^2
           -2.893(B-V)^3
\end{equation}
to make the correction
\begin{equation}
\rphk=R_{\mbox{\scriptsize HK}}-R_{\mbox{\scriptsize phot}}.
\end{equation}

From these \rphk values one can derive rotation periods from the
empirical fits of Noyes et al.:
\begin{equation}
\log{(P_{\mbox{\scriptsize rot}}/\tau)}=0.324
           -0.400 \log{R_5}
           -0.283 (\log{R_5})^2
	   -1.325 (\log{R_5})^3
\end{equation}
where $R_5$ is defined as $\rphk \times 10^5$ and $\tau$ is the
convective turnover time:
\begin{equation}\label{rotp}
\log{\tau} =  \left\{  \begin{array}{ll}
	           1.362
		  -0.166 x
                  +0.025 x^2
	          -5.323 x^3 & x > 0 \\
                  
	           1.362 
	          -0.14  x   & x < 0 \\
                      \end{array}   \right.
\end{equation}
where $x=1-(B-V)$ and the ratio of mixing length to scale height is 1.9.  Finally, we can calculate ages \citep[][cited in
\citet{Henry1996}]{Donahue1993}:
\begin{equation}
\log{t}=10.725
      -1.334 R_5
      +0.4085 R_5^2
      -0.0522 R_5^3
\end{equation}
where t is the stellar age in years.  The age calibration is certainly
invalid in the T Tauri regime, therefore in Table~\ref{punchline} for stars so
active that this relation yields unreasonably low ages of
$\log{\mbox{(Age}/\mbox{yr)}}<7$ we simply quote ``$< 7$''.

\citet{Noyes1984} report that the rms in the calibration for Equation~\ref{rotp} is 0.08 dex. \citet{Henry1996} notes that the age relation
yields ages such that in 15 of 22 binaries where it has been tested
the ages differ by less than 0.5 Gyr.  On the other hand, during the
solar cycle the sun's age as calculated by the relation varies from
2.2 to 8.0 Gyr.

For all three of these values, \rphk, $P_{\mbox{\scriptsize{rot}}}$,
and age, the reader should keep in mind that the {\em mean} S value of
a star is the dispositive quantity, and that many program stars have
been observed only once or twice.  Only for stars with many years of
observation can we claim good knowledge of a mean value of $S$.

\subsection{Tables of Measurements\label{tablelabel}}
We present two tables of S values here.  Table~\ref{seasonsample}
contains \sdiff\ values from all of our Keck observations.  
The first three columns of these tables identify the star observed by HD number,
Hipparcos number, and an ``other'' designation such as Gliese, HR, or
SAO number.  For some stars we have added binary component letters 'A' and 'B' to HD
numbers of the brighter and fainter components, respectively, for
uniqueness even when these qualifiers do not appear in 
the HD catalog.  The fourth column specifies the Julian Date of the
observation, and the fifth column lists the differential S value (on
the absolute Mount Wilson scale, as described in \S~\ref{differential}) for that observation.  The final column
contains alternate names for some stars and coordinates for stars for
which only one catalog name is given.  

Table~\ref{punchline} contains our median, final S value for each
star in our sample, which we refer to as the ``grand S''.  To remove
the effects of highly uneven sampling of stars which vary in 
activity, we took the median S values in 30-day bins and used the
median value of those medians.  For simplicity, and
to reduce the chance of breaking up observations taken within days of
each other, these bins are not adjacent but are rather defined
algorithmically such that some observation always lies an the
beginning of a 30-day interval.  For instance if observations occur on
days 1, 25, 62, 63, 90, 91 and 99, then the bins would be from days 1
through 30, days 62 through 91, and days 99 through 128.

To calculate grand S values we use only Keck and Dewar 6 spectra,
if possible.  For stars with only (less precise) Dewar 8 data, those are used and so
noted.  We combined all Dewar 6 and Keck observations for each of our
stars and calculated 30-day medians.  The median of all of these 30-day S
values we call the ``grand S'' value of the star, and the standard
deviation of the differential S values is quoted as a fractional
uncertainty, $\sigma_{S_{\mbox{\tiny diff}}} / S$. 

Table~\ref{punchline} contains 15 columns.  The first three identify
each star with the conventions described above for
Table~\ref{seasonsample}.  The fourth and fifth columns 
list $B-V$ and $M_V$ as reported by the Hipparcos catalog
\citep{Hipparcos} or, if unavailable, by SIMBAD. 
The sixth and seventh columns list the Julian dates of the first
and last observations used in calculating the grand S value.  The 
eighth column lists the total number of observations, and the ninth column lists the
number of monthly bins used to calculate the grand S value.  The
tenth column contains the grand S value as described above.  The
eleventh column contains the fractional standard deviation of the
differential S values for that star.  Since the precision of \sdiff\
is 1\%, entries in this column near 1\% represent stars
with without significant detected activity variations.

The next three columns list the quantities derived from these
measurements and target notes, $\log{\rphk}$,
$\log{(\mbox{age}/\mbox{yr})}$, and rotation period in days, as
described in \S~\ref{derived}.  The final column contains target
notes.  The note ``d8'' refers to an entry based solely on
Dewar 8 data, for which an the uncertainty of each measurement
contributing to the grand S is around 10\%.  All other entries are
based on Keck and Dewar 6 data
which have per-measurement uncertainties of around 6\%.  Again,
alternate names for stars are noted, and J2000 coordinates are given for stars with only one catalog name.

\subsection{Target Notes\label{targetnotes}}
\subsubsection{The Sun}
Our sample contains 5 observations of the asteroid Vesta obtained in
1997, about one
year after solar minimum.  We include these at the top of Table~\ref{punchline} under the name ``Sol''.  The S value of 0.167 is consistent
with solar minimum \citep{Baliunas1995}.  

\subsubsection{HD 531 A \& B}
The binary system of HD 531 consists of two stars of similar
colors and magnitudes separated by 5 \arcsec.  Since there seems to be
confusion in SIMBAD regarding the properties of and nomenclature for these stars, we
have deemed the eastern object HD 531B and the western one HD 531A,
and we have not listed colors for these objects.

\subsubsection{HR 3309}
HR 3309 (= HD 71148), which appears as a boxed cross in
Figure~\ref{keckcalib}, has a grand S value of 0.17, which is highly discrepant 
from the mean Duncan et al. value of 1.57.  The higher value is
inconsistent with the Hipparcos $B-V$ value of 0.67 for this star.
\citet{Soderblom1985}, in his analysis of the Mount Wilson S values,
quotes a mean S value of 0.169, consistent with our value, and the Mount
Wilson project archives confirm that the Duncan et al. value is a
transcription error (S. Baliunas, private communication).

\subsubsection{HD 137778}
HD 137778 (= GJ 586B), which appears as a boxed asterisk in
Figure~\ref{keckcalib}, has a grand S value of 0.57, significantly higher
than the mean Duncan et al. value of 0.16.  Both values are plausible
with its Hipparcos $B-V$ color of 0.87, but the higher value would imply a very young age.  \citet{Strassmeier2000} report a
``$R_{\mbox{\scriptsize HK}}$'' value of $7 \times 10^{-5}$
which apparently corresponds to the \rphk\ values calculated here.
Strassmeier's  $R_{\mbox{\scriptsize HK}}$ would imply $S = 0.875$ for this star, which is
extremely high for a star of this color, and would imply extraordinary activity.

This star is in the ROSAT All-Sky Survey bright source catalog with a flux of $\sim5 \times 10^{13}$ erg/$\mbox{cm}^2$/s.  With a parallax of 48 mas, this corresponds to $L_{\mbox{\scriptsize X}}/L_{\mbox{\scriptsize Bol}} \sim10^{-5}$, consistent with a very active, young star.

These discrepancies are puzzling, although it appears that most
measurements imply significant activity.  Perhaps this star has simply become
significantly more active since the Mount Wilson data were taken.
Observations by the Mount Wilson project between May 1995 and March
2001 give a mean S value of 0.64, much closer to the value reported here
(S. Baliunas, private communication).  The Strassmeier value may
represent an extraordinary event or may be calibrated differently than
values derived from S values.
 
\subsubsection{HD 58830}
Based upon its spectrum, HD 58830 (= GJ 9233), which appears as a
boxed triangle in Figure~\ref{keckcalib}, appears not to be a
main-sequence star at all, but an F5-F8 giant, although SIMBAD lists it as
a G0 star.  Our S value of 0.20 is significantly different from the
Duncan et al. value of 0.54.  Observations by the Mount Wilson project
between November 1997 and November 1998 give a mean S value of 0.21,
consistent with our observations (S. Baliunas, private communication).

\section{Conclusions}
We have measured chromospheric
activity as S values from over 15,000 archival spectra taken over
the course  of the California and Carnegie Planet Search Program.
These spectra were taken with the HIRES spectrograph at Keck
Observatory and the two detectors at the Hamilton spectrograph at Lick
observatory and contain both precision velocity information and the
\caii\ lines from which S values were derived.

Extraction of activity measurements from the Keck spectra was
successful, with over 95\% of all Keck spectra used yielding useful S
values.  The Lick spectra were more problematic: only $\sim 50\%$ of the spectra
proved useful due to low signal to noise ratios and poor extraction in
this very blue region by the 
automated extraction pipeline.  Nonetheless, the 1400 good Lick
measurements and the 14514 good Keck measurements combine to give a
record of the chromospheric activity for over 1000 late-type 
main-sequence stars.  Analysis of the measured activity level of
$\tau$ Ceti demonstrates a typical per-observation random error
of 6\% at Keck and 6.5\% and 10.5\% for the two detectors at Lick.  

We combined the Keck data with the good Lick data to 
create median S values in 30-day bins.  We have generated ``grand S''
by taking the median of
these monthly values.  These grand S's represent median activity levels
for our program stars for the periods we observed them.  For stars
with $0.5<B-V<0.9$, there are well-calibrated 
relationships between mean activity level, age, and rotation period,
allowing the determination of those quantities for our stellar
sample.  We present our grand S values and derived ages and rotation periods in
Table~\ref{punchline}.  We also have measured differential S values
for each stars which are more precise.  We present these data electronically as Table~\ref{seasonsample}.

For each star, these measurements of activity, $S$ and \rphk\  represent median
activity levels over the duration of the observations, which may be
significantly shorter than a stellar activity cycle.  This represents
a source of error when deriving stellar properties such as age from
these measurements.

The differential S values, \sdiff, measured here are much
more precise than $S$ but no more accurate:
we scaled them to the median of the S values, so while they are
excellent measurements of temporal variations in activity for a given
star they contain no additional information regarding differences in
the overall activity among stars.  

We have also measured metallicities for many of these stars and in
a future work \citep{Wright2004} will present an analysis of these
same stars for stellar characteristics, chromospheric periodicities,
and evolutionary status, including the effects of metallicity from
\citet{Valenti2004} and \citet{Fischer2004}.

\acknowledgments
The authors are indebted to Debra Fischer for her work with the Lick
Observatory Planet Search, particularly in regards to gathering \caii\
data, and for donating her time and expertise regarding the Hamilton
spectrograph and its data products. 

The authors wish to recognize and acknowledge the
very significant cultural role and reverence that the summit of Mauna
Kea has always had within the indigenous Hawaiian community.  We are
most fortunate to have the opportunity to conduct observations from
this mountain.

The authors also thank the many observers who helped
gather the data herein, and the referee, Sallie Baliunas, for her
insightful and constructive report.

This research has made use of the SIMBAD database, operated at CDS,
Strasbourg, France.  

This research has made use of NASA's Astrophysics Data System Bibliographic Services.
\begin{deluxetable}{lllcccl}
\tablecolumns{9}
\tablewidth{0pc} 
\tablecaption{\sdiff\ for Planet Search Program stars\label{seasonsample}}
\tablehead{ 
\colhead{HD}& \colhead{HIP} & \colhead{Other} & \colhead{Date}   &
\colhead{\sdiff} & \colhead{notes}\\ 
\colhead{} & \colhead{} & \colhead{} & \colhead{JD-2400000} &
\colhead{}}
\startdata 
 &  & Sol & 50667.07115 & 0.168 & Vesta\\
 &  & Sol & 50667.07810 & 0.165 & Vesta\\
 &  & Sol & 50667.08616 & 0.167 & Vesta\\
 &  & Sol & 50667.09427 & 0.165 & Vesta\\
 &  & Sol & 50667.10113 & 0.166 & Vesta\\
      225261  &          400 & GJ 3003 & 50366.92005 & 0.176 & \\
      225261  &          400 & GJ 3003 & 50667.02786 & 0.172 & \\
      225261  &          400 & GJ 3003 & 50689.03329 & 0.177 & \\
      225261  &          400 & GJ 3003 & 50716.03115 & 0.171 & \\
      225261  &          400 & GJ 3003 & 51010.02209 & 0.170 & \\
      225261  &          400 & GJ 3003 & 51071.00666 & 0.175 & \\
      225261  &          400 & GJ 3003 & 51072.00845 & 0.170 & \\
      225261  &          400 & GJ 3003 & 51072.91727 & 0.177 & \\
      225261  &          400 & GJ 3003 & 51173.80237 & 0.173 & \\
      225261  &          400 & GJ 3003 & 51368.01875 & 0.175 & \\
\enddata

\tablecomments{[The complete version of this table is in the electronic edition of
the Journal.  The printed edition contains only a sample.]}
\end{deluxetable}

\begin{deluxetable}{lrlcrcccccrrrrl} 
\rotate
\tabletypesize{\scriptsize}
\setlength{\tabcolsep}{0.05in}
\tablecolumns{15} 
\tablewidth{0pc} 
\tablecaption{Measured and Derived Quantities for all
Stars\label{punchline}}
\tablehead{ 
\colhead{HD}& \colhead{HIP}& \colhead{Other} & \colhead{B-V} &
\colhead{$M_V$} & \colhead{Begin}   &  \colhead{End} &
\colhead{\# Obs} & \colhead{\# Month} & \colhead{Grand S}
& \colhead{$\sigma_{S_{\mbox{\tiny diff}}} / S$} &
\colhead{log \rphk} & \colhead{P$_{\mbox{\scriptsize rot}}$} &
\colhead{log (Age/yr)} & \colhead{Note} \\ 
\colhead{} & \colhead{} & \colhead{} & \colhead{} & \colhead{} &
\multicolumn{2}{c}{JD-2400000} & \colhead{} & \colhead{Bins} & \colhead{} &
\colhead{\%} & \colhead{} & \colhead{(days)} & \colhead{} & \colhead{}}
\startdata 
 &  & Sol & 0.656 &   4.83 & 50667 & 50667 &     5 &     1 & 0.166 &   0.70 & -4.96 &   25. &  9.69 & Vesta; See \S~\ref{targetnotes}\\
      224983  &          184 &  & 0.888 &   5.85 & 52833 & 52833 &     1 &     1 & 0.177 & \nodata & -5.01 &   46. &  9.76 & \\
      225261  &          400 & GJ 3003 & 0.755 &   5.78 & 50367 & 52836 &    32 &    18 & 0.174 &   1.36 & -4.96 &   36. &  9.68 & \\
 &          428 & GJ 2F & 1.472 &   9.65 & 51757 & 52829 &    16 &    10 & 1.788 &   9.44 & \nodata\tablenotemark{1} & \nodata\tablenotemark{1} & \nodata\tablenotemark{1} & \\
      225213  &          439 & GJ 1 & 1.462 &  10.36 & 52601 & 52832 &     2 &     2 & 0.451 &  16.21 & \nodata\tablenotemark{1} & \nodata\tablenotemark{1} & \nodata\tablenotemark{1} & \\
          38 A &          473 & GJ 4A & 1.440 &   8.66 & 51757 & 52829 &    12 &     9 & 1.647 &   4.28 & \nodata\tablenotemark{1} & \nodata\tablenotemark{1} & \nodata\tablenotemark{1} & \\
          38 B &  & GJ 4B & 1.410 & \nodata & 51757 & 52829 &    19 &    10 & 1.820 &   6.12 & \nodata\tablenotemark{1} & \nodata\tablenotemark{1} & \nodata\tablenotemark{1} & \\
         105  &          490 & SAO 214961 & 0.595 &   4.49 & 52516 & 52833 &     7 &     3 & 0.337 &   2.92 & -4.41 &    5. &  8.63 & \\
\enddata

\tablecomments{$B-V$ and $M_V$ values are from Hipparcos \citep{Hipparcos}, where available.  \textit{Begin} and \textit{End} are the dates of the
first and last observations used to compute the grand S. \textit{\# Month Bins}
denotes total number of time bins used to determine the grand S as
defined in \S~\ref{season}. \textit{\# Obs} denotes the total number of
observations of the target included in grand S.  $\sigma_{S_{\mbox{\tiny diff}}} / S$ is the
fractional standard deviaton of the differential S values for a star
(for those stars with Keck data) and measures the star's intrinsic variability.  The
overall uncertainty in the grand S varies from star to star, but we
conservatively estimate it to be around 13\%, in general (see \S~\ref{errors}).
\textit{log (Age/yr)} values less than 7 are entered as ``$< 7$''.
Coordinates in the \textit{Note} column are J2000 Hipparcos
coordinates. Other entries in the \textit{Note} column are explained in \S~\ref{tablelabel}}

\tablenotetext{1}{The chromospheric activity relation used to
derive this quantity is only calibrated for stars with $0.44 < B-V <
0.9$.}

\end{deluxetable}


\begin{thebibliography}{}
\bibitem[Baliunas et al.(1995)]{Baliunas1995} Baliunas, S.~L.~et al.\ 1995, \apj, 438, 269 
\bibitem[Baliunas et al.(1998)]{Baliunas1998} Baliunas, S.~L., Donahue, R.~A., Soon, W., \& Henry, G.~W.\ 1998, ASP Conf.~Ser.~154: Cool Stars, Stellar Systems, and the Sun, 10, 153 
\bibitem[Butler et al.(1996)]{Butler1996} Butler, R.~P., Marcy, G.~W., Williams, E., McCarthy, C., Dosanjh, P., \& Vogt, S.~S.\ 1996, \pasp, 108, 500 
\bibitem[Butler et al.(2003)]{Butler2003} Butler, R.~P., Marcy, G.~W.,
Vogt, S.~S., Fischer, D.~A., Henry, G.~W., Laughlin, G., \& Wright, J.~T.\ 2003, \apj, 582, 455 
\bibitem[Cumming, Marcy, \& Butler(1999)]{Cumming1999} Cumming, A., Marcy, G.~W., \& Butler, R.~P.\ 1999, \apj, 526, 890 
\bibitem[Donahue (1993)]{Donahue1993} Donahue, R.~A., 1993 Ph.D. thesis, New Mexico State University
\bibitem[Duncan et al.(1991)]{Duncan1991} Duncan, D.~K.~et al.\ 1991, \apjs, 76, 383 
\bibitem[Filippenko(1982)]{Filippenko1982} Filippenko, A.~V.\ 1982, \pasp, 94, 715
\bibitem[Fischer \& Valenti (2004)]{Fischer2004} Fischer, D.~A. \& Valenti, J.~A.\ 2004, in prep.
\bibitem[Hartmann et al.(1984)]{Hartmann1984} Hartmann, L., Soderblom, D.~R., Noyes, R.~W., Burnham, N., \& Vaughan, A.~H.\ 1984, \apj, 276,254 
\bibitem[Henry, Donahue, \& Baliunas(2002)]{Henry2002} Henry, G.~W., Donahue, R.~A., \& Baliunas, S.~L.\ 2002, \apjl, 577, L111 
\bibitem[Henry et al.(1996)]{Henry1996} Henry, T.~J., Soderblom, D.~R., Donahue, R.~A., \& Baliunas, S.~L.\ 1996, \aj, 111, 439 
\bibitem[Horne(1986)]{Horne1986} Horne, K.\ 1986, \pasp, 98, 609 
\bibitem[Middelkoop(1982)]{Middelkoop1982} Middelkoop, F.\ 1982, \aap, 107, 31 
\bibitem[Nidever et al.(2002)]{Nidever2002} Nidever, D.~L., Marcy, G.~W., Butler, R.~P., Fischer, D.~A., \& Vogt, S.~S.\ 2002, \apjs, 141, 503 
\bibitem[Noyes et al.(1984)]{Noyes1984} Noyes, R.~W., Hartmann, L.~W., Baliunas, S.~L., Duncan, D.~K., \& Vaughan, A.~H.\ 1984, \apj, 279, 763 
\bibitem[Perryman et al.(1997)]{Hipparcos} Perryman, M.~A.~C.~et al.\ 1997, \aap, 323, L49 
\bibitem[Queloz et al.(2001)]{Queloz2001} Queloz, D.~et al.\ 2001, \aap, 379, 279 
\bibitem[Saar \& Fischer(2000)]{Saar2000} Saar, S.~H.~\& Fischer, D.\ 2000, \apjl, 534, L105 
\bibitem[Santos et al.(2000)]{Santos2000} Santos, N.~C., Mayor, M., Naef, D., Pepe, F., Queloz, D., Udry, S., \& Blecha, A.\ 2000, \aap, 361, 265 
\bibitem[Santos et al.(2003)]{Santos2003} Santos, N.~C. et al.\ 2003, astro-ph/0305434
\bibitem[Shkolnik, Walker, \& Bohlender(2003)]{Shkolnik2003} Shkolnik, E., Walker, G.~A.~H., \& Bohlender, D.~A.\ 2003, \apj, 597, 1092 
\bibitem[Soderblom(1985)]{Soderblom1985} Soderblom, D.~R.\ 1985, \aj, 90, 2103 
\bibitem[Strassmeier et al.(2000)]{Strassmeier2000} Strassmeier, K., Washuettl, A., Granzer, T., Scheck, M., \& Weber, M.\ 2000, \aaps, 142, 275 
\bibitem[Tinney et al.(2002)]{Tinney2002} Tinney, C.~G., McCarthy, C., Jones, H.~R.~A., Butler, R.~P., Carter, B.~D., Marcy, G.~W., \& Penny, A.~J.\ 2002, \mnras, 332, 759 
\bibitem[Valenti \& Fischer (2004)]{Valenti2004} Valenti, J.~A. \& Fischer, D.~A.\ 2004, in prep.
\bibitem[Vaughan, Preston, \& Wilson(1978)]{Vaughan1978} Vaughan, A.~H., Preston, G.~W., \& Wilson, O.~C.\ 1978, \pasp, 90, 267 
\bibitem[Vogt et al.(1994)]{Vogt1994} Vogt, S.~S., et al. 1994, Proc. Soc. Photo-Opt. Instr. Eng., 2198, 362
\bibitem[Vogt(1987)]{Vogt1987} Vogt, S.~S.\ 1987, \pasp, 99, 1214 
\bibitem[Wilson(1968)]{Wilson1968} Wilson, O.~C.\ 1968, \apj, 153, 221 
\bibitem[Wright(2004)]{Wright2004} Wright, J.~T.\ 2004, in prep
\end{thebibliography}
\end{document}